# RETHINKING THE PALEOPROTEROZOIC GREAT OXIDATION EVENT: A BIOLOGICAL PERSPECTIVE

John W. Grula

*Observatories of the Carnegie Institution for Science*
*813 Santa Barbara Street*
*Pasadena, CA 91101 USA*
*jgrula@obs.carnegiescience.edu*

## ABSTRACT

Competing geophysical/geochemical hypotheses for how Earth's surface became oxygenated – organic carbon burial, hydrogen escape to space, and changes in the redox state of volcanic gases – are examined and a more biologically-based hypothesis is offered in response. It is argued that compared to the modern oxygenated world, organic carbon burial is of minor importance to the accumulation of oxygen in a mainly anoxic world where aerobic respiration is not globally significant. Thus, for the Paleoproterozoic Great Oxidation Event (GOE) ~ 2.4 Gyr ago, an increasing flux of $O_2$ due to its production by an expanding population of cyanobacteria is parameterized as the primary source of $O_2$. Various factors would have constrained cyanobacterial proliferation and $O_2$ production during most of the Archean and therefore a long delay between the appearance of cyanobacteria and oxygenation of the atmosphere is to be expected. Destruction of $O_2$ via $CH_4$ oxidation in the atmosphere was a major $O_2$ sink during the Archean, and the GOE is explained to a significant extent by a large decline in the methanogen population and corresponding $CH_4$ flux which, in turn, was caused primarily by partial oxygenation of the surface ocean. The partially oxygenated state of these waters also made it possible for an aerobic methanotroph population to become established. This further contributed to the large reduction in the $CH_4$ flux to the atmosphere by increasing the consumption of $CH_4$ diffusing upwards from the deeper anoxic depths of the water column as well as any $CH_4$ still being produced in the upper layer. The reduction in the $CH_4$ flux lowered the $CH_4$ oxidation sink for $O_2$ at about the same time the metamorphic and volcanic gas sinks for $O_2$ also declined. As the $O_2$ source increased from an expanding population of cyanobacteria – triggered by a burst of continent formation ~ 2.7-2.4 Gyr ago – the atmosphere flipped and became permanently oxygenated.

*Subject headings*: oxygen; atmospheric oxygenation; oceanic oxygenation; cyanobacteria; methanogens; methanotrophs

# 1. INTRODUCTION

The astrobiological implications of detecting free oxygen ($O_2$ and/or $O_3$) in the atmosphere of an Earth-like exoplanet are highly significant, and under most circumstances such a discovery would be considered strong evidence for the existence of life on that planet (Sagan *et al*., 1993; Kasting, 2010; Léger *et al*., 2011).  Meanwhile, the causes of Earth's oxygenation and exactly how its oceans and atmosphere came to have abundant $O_2$ continue to be the subject of intense investigation and much debate.  While there is general agreement that a complex set of geophysical, geochemical, and biological processes were probably involved, geophysical and geochemical processes continue to dominate thinking about this subject (e.g., see Kump, 2008 and Holland, 2009).  While at least some of these factors were no doubt important in oxygen's rise, biological factors perhaps similar in importance need to be given full consideration.

Among Earth scientists there are currently three basic schools of thought about which geophysical/geochemical process was most important in causing the first Great Oxidation Event (GOE; Holland, 2002) approximately 2.4-2.2 Gyr ago during the Paleoproterozoic.  A long standing position, recently restated by Falkowski and Isozaki (2008), is that the GOE was caused by an increase in the burial of organic carbon: "without the burial of organic matter in rocks, there would be very little free $O_2$ in the atmosphere."  A second view, championed primarily by David Catling and coworkers, is the most important mechanism that ultimately caused the GOE was the oxidation of Earth's crust by enhanced hydrogen escape into space as a result of ultraviolet photolysis of abundant biogenic methane in the upper atmosphere (Hunten & Donahue, 1976; Catling *et al*., 2001; Catling and Claire, 2005; Claire *et al*., 2006).  The third position places greatest emphasis on how geochemical sinks for oxygen in the form of reduced volcanic and metamorphic gases may have decreased over time as the oxidation state of these gases increased (e.g., Holland, 2002 and Holland, 2009).  Recent variations on this theme include the idea that an increase in subaerial volcanism around 2.5 Gyr ago diminished the sink for oxygen because the gases emanating from such volcanoes were less reducing than the gases released from submarine volcanoes (Kump and Barley, 2007; Gaillard *et al*., 2011), and the proposal that oxygenation occurred because the $CO_2/H_2O$ and $SO_2/H_2O$ ratios of volcanic gases increased over time (Holland, 2009).  Catling and coworkers have connected the second school of thought with the third by arguing that as hydrogen escape drove oxidation of the lithosphere this decreased the flux of reducing metamorphic gases derived from the crust (Catling *et al*., 2001; Claire *et al*., 2006).  However, hydrogen escape apparently was not a factor in the shift to less reducing mantle-derived gases from subaerial volcanoes (Holland, 2002; Sleep, 2005; Claire *et al*., 2006).  Instead, the change in the redox state of these gases is proposed to have resulted from a major tectonic event of continental stabilization at the Archean/Proterozoic transition that increased the proportion of subaerial volcanism to submarine volcanism, and as a result oxidized volcanic gases such as $H_2O$, $CO_2$, and $SO_2$ became more dominant (Kump and Barley, 2007; Gaillard *et al*., 2011).

In addition to arguing for the merits of their respective hypotheses, the various advocates have often expressed doubts regarding the importance of the other explanations for oxygen's rise. For example, Catling, Kasting, Kump, and coworkers have expressed doubts that an increase in organic carbon burial was the cause of the GOE (Kasting, 1993; Catling and Claire, 2005; Kump and Barley, 2007), and two recent models for the GOE presented by Zahnle *et al*. (2006) and Holland (2009) do not discuss organic carbon burial at all. On the other hand, Falkowski and Isozaki (2008) have argued hydrogen escape and changes in the redox state of volcanic gases are oversimplifications. At the same time, Kump and Barley (2007) make no mention of hydrogen escape and explain the Paleoproterozoic rise in atmospheric $O_2$ in terms of an increase in the oxidation state of volcanic gases. Holland's most recent proposal maintains the GOE resulted from an increase in the $CO_2/H_2O$ and $SO_2/H_2O$ ratios of volcanic gases while the $H_2/H_2O$ ratio of these gases remained constant (Holland, 2009). He also states "there is no direct evidence to support" the hydrogen escape hypothesis and "there is some evidence to the contrary" as to whether or not enhanced hydrogen escape resulted in progressive oxidation of the continental crust (Holland, 2009). Suffice it to say that the disagreements among Earth scientists on this subject are substantial, and a consensus has yet to emerge. Indeed, according to Catling and Kasting (2007), "There is still no consensus about why atmospheric $O_2$ levels increased in the manner indicated by the geologic record." As such, new viewpoints should be welcomed into the discussion.

Here I first examine why a substantial delay between the appearance of cyanobacteria and oxygenation of the atmosphere is to be expected. Then some difficulties with the organic carbon burial hypothesis for explaining the Paleoproterozoic GOE are analyzed, and in this context the evolution and radiation of aerobic respiration are discussed. Implications for the GOE of the diverse metabolic paths for organic matter in the oceans and the existence of recalcitrant dissolved organic matter (RDOM) are also examined. Finally, I put forward a more biologically-based hypothesis for explaining the GOE that acknowledges the importance of at least some geophysical/geochemical processes while also arguing that microbial population dynamics, the physiological status of certain microbes, and other biological processes were perhaps of equal or greater importance.

## 2. A BIOLOGICAL PERSPECTIVE ON SOME OXYGENATION CONUNDRUMS

### 2.1. *A substantial delay between the appearance of cyanobacteria and atmospheric oxygenation is to be expected*

Many investigators of oxygen's rise on Earth have noted there was an apparent delay of at least several hundreds of millions of years between the first appearance of oxygenic photosynthesis by cyanobacteria about 2.7 Gyr ago and possibly earlier, and the first permanent accumulation of small amounts of atmospheric oxygen about 2.4 Gyr ago (e.g., see Catling *et al*., 2001; Bekker *et al*., 2004; Goldblatt *et al.,* 2006; Kump and Barley, 2007; Lyons 2007; Catling and Kasting, 2007). While a recent report has called into question some of the biomarker evidence for the existence of oxygen-producing cyanobacteria 2.7 Gyr ago (Rasmussen *et al.*, 2008), other investigators have maintained that various microbial fossil signatures as well as geological evidence in the form of hematite deposits and thick, widespread kerogenous shales still provide good reason to think cyanobacteria were probably present 2.7 Gyr ago or earlier (Buick, 2008; Fischer, 2008; Hoashi *et al.*, 2009; Waldbauer *et al.*, 2009). Isotopically light bulk kerogens dated at 2.7 Gyr ago have also been suggested to require oxygenic photosynthesis (Hayes, 1983; Hayes, 1994), and the same applies to enrichments of $^{53}$Cr in 2.7 Gyr-old iron formations (Frei *et al.*, 2009; Lyons and Reinhard, 2009). Bracketing this debate are the highly divergent views that cyanobacteria did not evolve until ~ 2.5-2.4 Gyr ago, but then quickly proliferated and triggered a major glaciation event ~ 2.3-2.2 Gyr ago (Kopp *et al.*, 2005), versus C and U-Th-Pb isotopic evidence that oxygenic photosynthesis (and therefore cyanobacteria or an ancestral form) evolved before 3.7 Gyr ago (Rosing and Frei, 2004). This controversy notwithstanding, here I will argue that a large delay between the appearance of cyanobacteria and oxygenation of the atmosphere is to be expected:

(1) The geochemical sinks for oxygen during the Archaean and early Proterozoic were vast and almost certainly much larger than they are now (Lowe, 1994; Holland, 2002; Claire *et al.*, 2006; Kump and Barley, 2007; Knoll, 2008), and therefore all of the oxygen initially produced by cyanobacteria would have been chemically bound by processes such as reaction with $Fe^{2+}$ in the oceans, combination with reduced volcanic and metamorphic gases, and crustal weathering (Canfield, 2005; Catling and Claire, 2005; Catling *et al.*, 2005; Kump and Barley, 2007). Because of the magnitude of these geochemical sinks, the oxygen produced by cyanobacteria could have been consumed for hundreds of millions of years until some combination of increased oxygen production and a decrease in these sinks finally made it possible for free oxygen to begin to accumulate (e.g., see Catling and Claire, 2005; Hayes and Waldbauer, 2006).

(2) Before the evolution of oxygenic photosynthesis, $CH_4$ production by methanogens using abundant $H_2$ and $CO_2$ from geological sources and acetate derived from anoxygenic photosynthesis probably made this gas an abundant constituent of the Archean atmosphere with a concentration over 1000X higher than it is in today's atmosphere (Kasting, 2005; Kharecha *et al.*, 2005; Kasting and Ono, 2006; Haqq-Misra *et al.,* 2008) and at least several orders of magnitude more abundant than $O_2$ (Zahnle *et al.*, 2006). In an anoxic atmosphere with no ozone shield, "oxygen is rapidly consumed in an ultraviolet-catalyzed reaction with biogenic methane" (Kasting, 2006). Therefore, the "mutual annihilation of $CH_4$ and $O_2$" (Claire *et al.*, 2006) would have also been a substantial $O_2$ sink that helped suppress its accumulation in the Archean atmosphere.

(3) With no oxygen available in the Archean atmosphere to form a stratospheric ozone shield capable of blocking most of the solar ultraviolet flux (which was probably more intense in the Archean than it is now because of the properties of the young sun [Canuto *et al.*, 1982; Zahnle and Walker, 1982; Walter and Barry, 1991; Cnossen *et al.*, 2007]), cyanobacteria and other life forms may have been severely challenged to cope with the direct effects of this potentially lethal radiation as well as the secondary effects of enhanced photooxidative damage (Garcia-Pichel, 1998). Atmospheric alternatives to an ozone shield, such as elemental sulfur vapor (Kasting and Chang, 1992) and organic hazes (Lovelock, 1988;

Pavlov et al., 2001; Wolf and Toon, 2010) have been proposed. However, whether an elemental sulfur vapor shield could have formed is by no means certain (Kasting and Chang, 1992) and early modeling indicated organic hazes may not have been very effective (Pavlov et al., 2001). On the other hand, more recent modeling of fractal organic hazes has shown that these particles, if they did indeed form in the Archean atmosphere as hypothesized, could have created an effective UV shield (Wolf and Toon, 2010).

If there was not an effective UV shield during the Archean, pelagic cyanobacteria of the open oceans probably took refuge in deeper waters (as much as 30 meters in depth?) where exposure to lethal UV would have been well attenuated (Kasting, 1987; Garcia-Pichel, 1998; Cockell, 2000). In the modern open ocean pelagic cyanobacteria such as Synechococcus, Trichodesmium, and Prochlorococcus (strictly speaking, a prochlorophyte) are the most abundant photosynthetic microorganisms and among the most important primary producers (Capone et al., 1997; Ferris and Palenik, 1998; Fuhrman and Campbell, 1998). Cyanobacteria similar to these would have been even more important contributors to primary productivity and $O_2$ production during the Archean because the total continental area and shallow coastal habitat were much smaller than they are now (approximately 5% of the present Precambrian continental crust existed ~ 3.1 Gyr ago, rising to about 60% by ~ 2.5 Gyr ago [Lowe 1994]) and open ocean covered a greater portion of Earth's surface than it does presently. Thus, the cyanobacteria living in shallow coastal waters probably made a small contribution to $O_2$ production during most of the Archean, and this did not increase until more substantial continental growth occurred at the very end of this eon. The UV-screening features which may have protected these shallow-water organisms, such as mat-forming habits and microbial biomineralization (Pierson, 1994; Phoenix et al., 2001), could not have formed in deep open waters to confer protection to pelagic cyanobacteria (Garcia-Pichel, 1998; Cockell, 2000).

If the surface UV radiation in the 200-300 nm range during the Archean was several orders of magnitude higher than current levels (Kasting, 1987; Cockell, 2000; Cnossen et al., 2007), even the existence of UV screens that may have had some effect in the open ocean -- such as dissolved reduced iron (Garcia-Pichel,1998) and nanophase iron oxides (Bishop et al., 2006) -- were probably not sufficient to allow pelagic cyanobacteria to exist close to the ocean surface. Therefore, this would have reduced the living space for these cyanobacteria and restricted the growth of their populations (Garcia-Pichel, 1998). Furthermore, the visible light reaching their narrow habitable zone deeper in the ocean would have been reduced in intensity (compounding visible light limitations already prevailing during the Archean due to the less luminous nature of the young sun [Newman and Rood, 1977; Gough, 1981]), and thus the rate of oxygenic photosynthesis would have been significantly reduced compared to that in modern oceans. As a result, this would have further constrained the growth of the cyanobacterial global population and their primary productivity. Their subdued rate of oxygen production would, in turn, have contributed to the delay in the rise of oxygen in the oceans and atmosphere (Garcia-Pichel, 1998).

(4) In addition to any limitations imposed by lethal UV, cyanobacterial proliferation during the Archaean was probably also constrained by nutrient availability, especially phosphorous (Bjerrum and Canfield, 2002; Papineau et al., 2007; Papineau et al., 2009;

Papineau, 2010).  When the flux of phosphorous into the oceans increased during the late Archean and earliest Paleoproterozoic, this probably triggered cyanobacterial blooms that led to the production of large amounts of $O_2$ (Papineau *et al.*, 2007; Papineau *et al.*, 2009; Papineau, 2010).  In addition, nitrogen may also have been in short supply, even if some or all cyanobacteria were able to biologically fix nitrogen (Kasting and Seifert, 2001; Navarro-Gonzalez *et al.*, 2001; Grula, 2005; Godfrey and Falkowski, 2009).  Lowe (1994) has further argued that the Archean ocean was probably highly stratified with little input of dissolved minerals from continents, and thus most of the surface layer may well have been a "nutrient-depleted biological desert" where cyanobacterial proliferation would have been very slow.

The combination of lethal UV and nutrient constraints could have greatly limited cyanobacterial population growth, and oxygen production, for many hundreds of millions of years.  In this context it is worth noting that the phytoplankton biomass in modern oceans can vary by a factor 100 depending on the natural availability of nutrients, light levels, and temperature (Doney, 2006).  Thus, "environmental oxygen levels could have remained low long after the origin of oxygenic photosynthesis if rates of cyanobacterial photosynthesis were limited" (Knoll, 2008) and "the currently dominant oxygenic photosynthesis existed in limited environments before it became dominant and did not immediately produce oxygen-rich air when it did" (Sleep, 2005).  Moreover, this means assertions such as "cyanobacteria evolved and radiated shortly before" triggering the Makganyene "snowball Earth" event ~ 2.3-2.2 Gyr ago (Kopp *et al.*, 2005), and "oxygenic photosynthesizers probably radiated quickly and became dominant players in the planetary ecosystem… as they are today" (Claire *et al.*, 2006) may need to be reconsidered.

(5)  Because most oxygenic photosynthesis during the Archean was conducted by marine cyanobacteria living in the ocean's upper layer (with perhaps some occurring in relatively small fresh water ecosystems [Blank and Sanchez-Baracaldo, 2010]), any oxygen that was not first bound by ferrous iron and other reduced chemicals in oceanic surface waters (Kasting, 1987; Kasting and Ono, 2006) would have remained dissolved for some period of time in so-called "oxygen oases" before it would start accumulating in the atmosphere (Kasting, 1992; Waldbauer *et al.*, 2011).  Thus, the oceans retained significant amounts of $O_2$ in solution and this, too, would have contributed to the long delay between the appearance of cyanobacteria and atmospheric oxygenation.  Indeed, recent data indicate that at least some regions of the surface ocean were oxygenated for at least 50 Myr and perhaps as long as 300 Myr before the atmosphere finally became permanently oxygenated to a low level during the GOE (Eigenbrode and Freeman, 2006; Holland, 2006; Kaufman *et al.*, 2007; Anbar *et al.*, 2007; Garvin *et al.*, 2009; Godfrey and Falkowski, 2009; Kendall *et al.*, 2010).  In addition, the transfer of $O_2$ from surface waters to the atmosphere and to anoxic waters at lower depths would have been impeded by chemical gradients (Kasting, 1992; Waldbauer *et al.*, 2011).  Such gradients would have slowed the rate of $O_2$ transfer from the surface ocean to the atmosphere, and only limited sectors of the atmosphere near oxygen oases would have initially experienced some degree of oxygenation (Kasting, 1992; Pavlov and Kasting, 2002; Brocks *et al.*, 2003; Haqq-Misra *et al.*, 2011).

## 2.2. *Organic carbon burial is of minor importance to the accumulation of oxygen in a mainly anoxic world where aerobic respiration is not globally significant*

The idea that organic carbon burial is the source of atmospheric oxygen is apparently traceable to a 1971 paper by Lee Van Valen (Kasting, 1993), and this notion has become widely accepted to be applicable across all of geologic time. According to Van Valen (1971), oxygenic photosynthesis produces stoichiometrically equal amounts of oxygen and reduced carbon (although the universality of this is now in doubt; see Behrenfeld *et al*., 2008, Suggett *et al*., 2009, and Zehr and Kudela, 2009), and because "Almost all the oxygen is eventually used to oxidize the reduced carbon. Most of this oxidation occurs in respiration – of animals, of decomposers, and of plants themselves." "The only net gain in oxygen equals the amount of reduced carbon buried before it is oxidized" (Van Valen, 1971). More recently, Falkowski and Isozaki (2008) restated the argument as follows: "The presence of $O_2$ in the atmosphere requires an imbalance between oxygenic photosynthesis and aerobic respiration on time scales of millions of years; hence, to generate an oxidized atmosphere, more organic matter must be buried than respired."

The problem with this argument is that when oxygenic photosynthesis by cyanobacteria first arose at least 2.7 Gyr ago (Waldbauer *et al*., 2009), the free $O_2$ content of Earth's surface was negligible (Catling and Claire, 2005). Thus the opposite process – aerobic respiration – would have been mainly confined to small oxygen oases in shallow costal oceans (Eigenbrode and Freeman, 2006; Zahnle *et al*., 2006; Waldbauer *et al*., 2009) that would have supported aerobic respiration primarily during diurnal periods when there was sufficient sunlight to power oxygenic photosynthesis (Sigalevich *et al*., 2000). Even in the case of oxygen oases, dissolved $O_2$ would have had to reach levels equivalent to 1% of the present atmospheric level (PAL) before aerobic respiration could become more energetic than anaerobic fermentation (Knoll and Holland, 1995; Goldblatt *et al*., 2006).

Therefore, on a global scale very little $O_2$ would have been consumed by aerobic respiration. As a result, before $O_2$ became universally plentiful "there is no reason to expect that respiration should so nearly cancel $O_2$ emission as it does today" (Zahnle *et al*., 2006). Accordingly, on the mainly anoxic Earth of the Archean and earliest Paleoproterozoic, aerobic respiration was at most a very small $O_2$ sink, and thus the *initial* rise in atmospheric $O_2 \sim 2.4$ Gyr ago must have been largely unrelated to the burial of organic carbon as a mechanism for preventing $O_2$ consumption by aerobic respiration. As stated by Kenneth Towe (1990), "without aerobic heterotrophic recycling of organic carbon, the burial of iron (and sulphur), *not the burial of organic matter*, [emphasis added] would be the dominant control over the oxygen transferred to the atmosphere." Thus, the argument that organic carbon burial controls the accumulation of $O_2$ on Earth's surface is applicable only to the more recent oxygenated/aerobic world and not to the mainly anoxic/anaerobic world that existed before the GOE $\sim 2.4$ Gyr ago.

To elaborate somewhat, when investigators write equations for oxygenic photosynthesis ($CO_2 + H_2O \rightarrow CH_2O + O_2$) and the reverse process, aerobic respiration and decay ($CH_2O + O_2 \rightarrow CO_2 + H_2O$), there may sometimes be a tendency to view these processes as simple, reversible chemical reactions as they might occur in a test tube. In reality, these reactions are catalyzed by a variety of organisms and the status of these organisms such as their abundance, distribution, and physiological state determines the rate at which these reactions occur. Depending on the status of the organisms in question, it might be possible for one reaction to proceed more rapidly than the other, and thus the products of that reaction will accumulate (unless they are consumed by other types of reactions). In this sense, therefore, oxygenic photosynthesis and aerobic respiration may not always have formed a "tight couple" (as argued, for example, by Knoll 2003, p. 103), and this would have been the case during the Archean when aerobic organisms were probably localized and slow-growing, whereas cyanobacteria were probably widespread and increasingly abundant during the last few hundred million years of this eon.

Of course, a crucial question here is: how closely did the evolution and radiation of aerobically respiring microorganisms track the evolution and radiation of cyanobacteria? Given the strong possibility that there was a substantial delay between the widespread radiation of cyanobacteria and the first accumulation of small amounts of free $O_2$ on a global scale, as well as the lethal effects on many Archean organisms of the $O_2$ produced by cyanobacteria (Blankenship et al., 2007), it would seem likely that the widespread radiation and proliferation of aerobically respiring microorganisms lagged considerably behind that of cyanobacteria.

### 2.3. *The evolution and proliferation of aerobic respiration*

Adequate amounts of $O_2$ must first be present before aerobic respiration can occur, and our understanding of the evolutionary history of this form of metabolism and microbial metabolism in general before ~ 2.5 Gyr ago is incomplete (Thamdrup, 2007; Waldbauer *et al.*, 2009). When free $O_2$ in the environment was not otherwise killing many of the organisms that first encountered it (Blankenship *et al.*, 2007), at what point did it become abundant enough for aerobic respiration to evolve and gain a significant energetic advantage over fermentation and other anaerobic metabolisms? Knoll & Holland (1995) have argued the first widespread radiation of aerobically respiring prokaryotes did not occur until *after* oxygen levels rose above 1% of the PAL (present atmospheric level) ~ 2 Gyr ago. One percent or $10^{-2}$ PAL is the oxygen level at which it has been generally thought that aerobic respiration by single-celled organisms starts to become possible (known as the "Pasteur Point") and energetically advantageous compared to fermentation (Berkner & Marshall, 1964; Knoll & Holland, 1995). More recently, Stolper *et al.* (2010) have provided evidence that the facultative aerobe *Escherichia coli K-12* can respire at $O_2$ levels as low as $10^{-7}$ PAL. However, the question remains as to exactly how much of an energetic advantage aerobic respiration confers at $10^{-7}$- $10^{-2}$ PAL compared to fermentation and other anaerobic metabolisms. Stolper *et al.* (2010) conducted their experiments under highly artificial conditions; normally *E. coli* switch to anaerobic metabolisms at $5 \times 10^{-3} - 2 \times 10^{-2}$ PAL (Becker *et al*., 1996). Furthermore, the mere presence of some amount of $O_2$ does not necessarily mean a particular aerobic process can occur or occur efficiently. For example, combustion requires an atmosphere that is at least 13% $O_2$ (Belcher & McElwain, 2008). If the energetic advantage of aerobic respiration at $10^{-7} - 10^{-2}$ PAL is small or nonexistent, this would have limited the selective advantage of this metabolism, slowed the proliferation of aerobically respiring microorganisms, and thus mitigated how rapidly and to what extent aerobic respiration became a significant $O_2$ sink.

In this context it is also pertinent that ocean scientists now set the limits to aerobic sea life in terms of a minimum dissolved $O_2$ concentration, usually $\sim 5$ μM. Below this concentration aerobic microbes inefficiently take up dissolved $O_2$ and start to use other electron acceptors (Brewer & Peltzer, 2009). (A dissolved $O_2$ concentration of 5 μM corresponds to an atmospheric $O_2$ mixing ratio of 0.4% by volume or 2% PAL at 25°C [Pavlov *et al*., 2003; Canfield, 2005].) In addition, laboratory experiments using pure cultures of aerobic methanotrophic bacteria have shown that as dissolved $O_2$ falls below 5 μM these bacteria start to exhibit a significant decline in their ability to oxidize $CH_4$ (Ren *et al*., 1997). Data such as these support the contention that once atmospheric $O_2$ reached 1% PAL (corresponding to $\sim 2.5$ μM dissolved $O_2$) this did not necessarily mean that aerobic respiration began conferring a large energetic and selective advantage over anaerobic metabolisms. Thus, exactly when and how rapidly aerobically respiring microorganisms proliferated and radiated remains highly uncertain.

Oxygen oases produced in cyanobacterial mats that formed in shallow coastal waters of the otherwise anoxic late Archean ocean have been invoked as the sites where aerobic respiration could have first evolved (Fischer, 1965; Brocks *et al*., 2003; Knoll, 2008). The small amount of aerobic respiration that could have occurred in these localized oxygenated regions of the coastal surface ocean would have created only a very small $O_2$ sink. Recent geological evidence for the existence of oxygen oases in the late Archean indicates they probably began in localized and isolated shallow-water habitats before gradually expanding to the photic zones of deeper waters between $\sim 2.7$ and $\sim 2.45$ Gyr ago, even as the atmosphere and deeper ocean remained anoxic (Brocks *et al*., 2003; Eigenbrode and Freeman, 2006; Eigenbrode *et al*., 2008; Godfrey and Falkowski, 2009; Kendall *et al*., 2010). However, other data may indicate that oxygen oases were present much earlier in the Archean (Hoashi *et al*., 2009). In any event, the existence of widespread and efficient aerobic respiration that would have created a globally significant oxygen sink (and thus greatly increased the importance of organic carbon burial as a mechanism for the further accumulation of oxygen) probably did not appear until *after* $\sim 2.4$-2.2 Gyr ago, when $O_2$ finally became permanently present in more than trace amounts in the atmosphere and surface ocean (Knoll, 2008).

### 2.4. *Other doubts about increases in organic carbon burial as the cause of the GOE*

Doubts about the contribution of either secular or pulsed increases of organic carbon burial to the GOE have been expressed by Catling and Claire (2005), Hayes and Waldbauer (2006), Kasting (2006), and Kump and Barley (2007) based on other grounds. For example, Catling and Claire (2005) have argued that a conservative interpretation of the carbon isotope record is not consistent with a secular increase in organic carbon burial rates from the early Archean forward in time through the Phanerozoic. Likewise, Kump and Barley (2007) assert that the lack of any secular trend rules-out a substantial increase in organic carbon burial between 2.5 and 2.4 Gyr ago, and thus the rise of $O_2$ at this time must have been due to a decline in the $O_2$ sinks.

On the other hand, a possible pulse in burial rates between 2.3-2.1 Gyr ago (which coincides with a very positive $\delta^{13}C_{carb}$ excursion) could not have caused a *permanent* increase in $O_2$, and because the proposed pulse "follows the rise of $O_2$, it cannot be its cause. Instead, perhaps the pulse is an effect of $O_2$" (Catling and Claire, 2005). This latter point is similar to Kasting's earlier observation that "it is not clear whether the change in organic carbon burial rate was a cause or a consequence of the rise in atmospheric $O_2$" (Kasting, 1993). In a similar critique, Hayes and Waldbauer (2006) have noted that the sequence of isotopic signals "is reversed from that expected," and if an organic carbon burial event caused the GOE and subsequent disappearance of the mass-independent fractionation of sulphur isotopes (MIF-S), then the loss of MIF-S "should not precede the first carbon-isotopic enrichments." Hayes and Waldbauer (2006) conclude that "levels of $O_2$ are only weakly and indirectly coupled" to rates of organic carbon burial and carbon isotopic signals.

Finally, another factor that makes the organic carbon burial argument problematic is the existence of plate tectonics means much organic carbon is never permanently subducted or stabilized in cratons, but is instead returned to the Earth's surface in various forms where it has another chance to combine with $O_2$ (as long as some $O_2$ is present), thus canceling $O_2$ gains (Catling *et al*., 2001; Hayes and Waldbauer, 2006; Jackson *et al.*, 2007; Falkowski and Isozaki, 2008).

2.5. *Diverse fates for organic carbon, recalcitrant dissolved organic matter, and the GOE*

The fate of the organic carbon produced by oxygenic photosynthesis in the mainly anoxic/anaerobic world that existed during most of the Archean was probably much more complex than has been assumed. This has significant implications for the GOE. For example, some geochemists' (e.g., Goldblatt *et al*., 2006) have argued that the major metabolic path for carbon before the GOE can be summarized as oxygenic photosynthesis followed by fermentation and methanogenesis and is represented by the following equation:

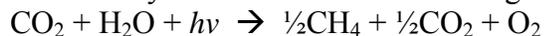
$$CO_2 + H_2O + h\nu \; \rightarrow \; \tfrac{1}{2}CH_4 + \tfrac{1}{2}CO_2 + O_2$$

While this would have produced a flux of $O_2$ and $CH_4$ to the atmosphere in a 2:1 stoichiometric ratio, any net $O_2$ gain was presumably nullified by atmospheric methane oxidation, which has been summarized by Catling and Kasting (2007) as follows:

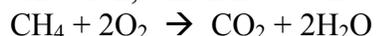
$$CH_4 + 2O_2 \; \rightarrow \; CO_2 + 2H_2O$$

In the effort to summarize various chemical reactions and the fate of their products, the complexity of biological processes such as fermentation, methanogenesis, and various poorly understood mechanisms that generate dissolved organic matter in the oceans (Ogawa *et al*., 2001; Jiao *et al*., 2010) should not be underestimated. In addition to $H_2$, $CO_2$, and acetate, other products of primary fermentation include propionate, butyrate, succinate, and alcohols. Of these, only $H_2$, $CO_2$, and acetate are directly convertible to $CH_4$ by methanogens, and thus other microorganisms – syntrophs – must be present to provide them with their needed substrates by breaking down more complex organic compounds (Madigan *et al*., 2003). Acetate, propionate, butyrate, succinate, alcohols, and $CO_2$ are all water-soluble compounds that could have accumulated to some extent in the partially oxygenated surface ocean of the late Archean – where cyanobacteria were photosynthesizing organic carbon and fermenters were also active – and thus deprived methanogens (which would have existed only in the anoxic lower ocean depths and sediments) of substrate that could have otherwise been used for $CH_4$ production if syntrophs were also present.

Even more importantly, it has recently been recognized that there is a vast reservoir of recalcitrant dissolved organic matter (RDOM) in the modern ocean, with an inventory of 624 Gt carbon that accounts for 95% of the total dissolved organic matter throughout the entire water column (Jiao *et al*., 2010). Under the anoxic conditions that prevailed in most of the water column during the late Archean and earliest Paleoproterozoic, the microbial carbon pump (MCP) responsible for the generation of RDOM was probably very effective, resulting in perhaps 500 times more RDOM than there is in today's oceans (Jiao *et al*., 2010; Stone, 2010). This huge and stable carbon reservoir (which probably left little, if any, isotopic signature in the geologic record) would have also deprived methanogens and their syntrophs of large amounts of organic matter generated by oxygenic photosynthesis.

In addition, because the organic matter generated by oxygenic photosynthesis in the surface ocean of the Archean also lacked an important mechanism found in the more recent ocean (since ~ 0.6 Gyr ago) for its efficient transport to the anoxic lower depths and sediments (repackaging into fast sinking faecal matter; Logan *et al*., 1995; Jiao *et al*., 2010), and to the extent that some of the organic carbon which *did* reach the sediments was permanently subducted away from methanogens and their syntrophs, these factors would have also deprived these organisms of substrate. The net effect of these processes was that the flux of $O_2$ and $CH_4$ to the atmosphere would have exceeded the 2:1 stoichiometric ratio expected if all the organic carbon produced by oxygenic photosynthesis became substrate for methanogenesis. Therefore, the $CH_4$ which *was* produced from organic carbon generated by oxygenic photosynthesis could not, by itself, have cancelled all $O_2$ gains via methane oxidation in the atmosphere. In addition, anaerobic methanotrophy of $CH_4$ in the oceans during the earliest Paleoproterozoic may have also contributed to making the ratio of the $O_2$ and $CH_4$ fluxes to the atmosphere greater than 2:1, and this would further "help $O_2$ win control of the redox state of the early atmosphere" (Catling *et al*., 2007).

## 3. THE PALEOPROTEROZOIC GREAT OXIDATION EVENT: MICROBES TAKE CENTER STAGE?

### 3.1. *Reformulating the Paleoproterozoic GOE*

In their paper "Biogeochemical modeling of the rise of atmospheric oxygen,"

Claire *et al.* (2006) formulate the accumulation of free oxygen on geological timescales as a function of $O_2$ sources ($F_{SOURCE}$) and sinks ($F_{SINK}$) as follows:

$$d/dt[O_2] = F_{SOURCE} - F_{SINK} = (F_B + F_E) - (F_V + F_M + F_W) \qquad (1)$$

In this equation $[O_2]$ is the total reservoir of atmospheric dioxygen. $F_{SOURCE}$ is divided into $F_B$, the flux of oxygen due to organic carbon burial, and $F_E$, the flux of oxygen to the Earth as a whole as a result of hydrogen escape into space. The $F_{SINK}$ is divided into $F_V$ and $F_M$, which represent reducing gases (for example, $H_2$, $H_2S$, CO, $CH_4$) from volcanic/hydrothermal and metamorphic/geothermal processes, respectively, while $F_W$ represents the oxygen sink caused by oxidative weathering of the crust.

I propose revising equation (1) as follows:

$$d/dt[O_2] = F_{SOURCE} - F_{SINK} = F_O - (F_G + F_{CH4} + F_{AR}) \qquad (2)$$

With respect to the $O_2$ sources, in this formulation $F_O$ is the flux of oxygen due directly to its production by cyanobacteria, and the size of this flux would have increased as the population of cyanobacteria increased. $F_B$ has been removed as it is made implicit by the addition of the $O_2$ sinks $F_{CH4}$ and $F_{AR}$ (discussed further below). $F_E$ has been removed because it is not a direct source of $O_2$; instead, it oxidizes the Earth's crust as a whole and in so doing lowers the $O_2$ sink due to the consumption of $O_2$ by reduced metamorphic gases, which is already represented by $F_M$ (this is discussed further in section 3.3 *Further analysis of $O_2$ sources and sinks*).

Regarding the individual $O_2$ sinks, $F_G$ represents the various geochemical sinks for oxygen including reaction with reduced volcanic and metamorphic gases, oxidative weathering of Earth's crust, and combination with iron and sulphur in the ocean. Therefore, $F_G$ essentially combines $F_V$, $F_M$, and $F_W$ in equation (1), plus it adds oceanic iron and sulphur sinks. $F_{CH4}$ is the sink due to the destruction of oxygen by biologically-produced methane via atmospheric methane oxidation (for more detailed discussions of the mutual destruction of $O_2$ and $CH_4$ via methane oxidation in the presence or absence of a stratospheric ozone layer, see Claire *et al.*, [2006] and Catling and Kasting, [2007]). Because atmospheric $CH_4$ was at least several orders of magnitude more abundant than $O_2$ during most of the Archaean (due mainly to a biological source, methanogenic archaea, which have probably existed on Earth since at least 3.46 Gyr ago [Ueno *et al.*, 2006]) and may well have had a mixing ratio of ~ 1000 ppmv or higher (Kasting, 2005; Kasting and Ono, 2006; Haqq-Misra *et al.*, 2008), biogenic $CH_4$ would have also been a significant $O_2$ sink during this period (Pavlov *et al.*, 2001; Pavlov and Kasting, 2002). This is because in the anoxic Archaean atmosphere without an ozone shield, $O_2$ would have been rapidly consumed via UV-catalyzed methane oxidation (Goldblatt *et al.*, 2006; Kasting, 2006). Finally, $F_{AR}$ represents the sink due to the consumption of $O_2$ by aerobic respiration and other aerobic metabolisms such as aerobic methanotrophy. While this sink was small before ~ 2.4 Gyr ago, it would increase in significance as free $O_2$ increased on Earth's surface and aerobically respiring organisms proliferated.

In their modeling of the rise of $O_2$ during the Paleoproterozoic GOE, Claire *et al.* (2006) discuss the mutual destruction of $CH_4$ and $O_2$ solely in terms of the atmospheric chemistry involved. They argue that because of a positive feedback due to decreasing destruction of $O_2$ by methane oxidation as stratospheric ozone forms, a geologically rapid switch occurred from an atmosphere where $CH_4$ is much more abundant than $O_2$ to one in which $O_2$ is more abundant than $CH_4$. Other modelers of the interaction between $CH_4$ and $O_2$ just before and during the GOE have obtained similar results (e. g., Goldblatt *et al.*, 2006).

In addition to such changes in atmospheric chemistry, another factor that would have made an important contribution to the rapid switch in the relative abundances of $CH_4$ and $O_2$ was the population dynamics of the microorganisms that produce and consume $CH_4$ – methanogens and methanotrophs – and the microorganisms that produce $O_2$, cyanobacteria. It is pertinent here that methanogens are obligate anaerobes which are extremely sensitive to oxygen and can only be cultured using strict anoxic techniques. Even brief exposure to a trace amount of oxygen is lethal to these organisms (Madigan *et al.*, 2003). Thus, an expansion of the cyanobacterial population in the late Archean and earliest Paleoproterozoic, which resulted in an increasingly oxygenated surface ocean (see next section), would have caused a significant decline in the anaerobic methanogen population in these waters while at the same time permitting an aerobic methanotroph population to become established. Ultimately the more complete oxygenation of the oceans and atmosphere by the end of the Proterozoic would relegate methanogens to relatively small refugia such as anoxic ocean sediments. On the modern Earth the remnant population of methanogens in these sediments makes a relatively small contribution to the total biogenic $CH_4$ flux (Madigan *et al.*, 2003), which is considerably smaller than that estimated for the late Archean (Kasting, 2005).

3.2. *The trigger for the GOE: massive continent formation and rifting, increased weathering and riverine delivery of nutrients and UV-absorbing chemicals to the oceans, followed by cyanobacterial proliferation*

During the late Archean and earliest Paleoproterozoic (2.7-2.4 Gyr ago) there was enormous growth in the lithospheric crust as major continental blocks formed for the first time and then broke-up (Aspler and Chiarenzelli, 1998; Lowe, 1994; Godderis and Veizer, 2000; Zhao *et al.*, 2004; Barley *et al.*, 2005; Dhuime *et al.*, 2011). This led to the creation of nearly 60% of the Precambrian continental crust that currently exists, up from the less than 5% that existed prior to 3.1 Gyr ago, an increase of over 10-fold (Lowe, 1994; Lowe and Tice, 2007). Supercontinent formation and the subsequent creation of intracratonic rift basins would have greatly improved the nutrient inventory of the oceans in two ways: First, the riverine input of dissolved minerals such as phosphate would have increased dramatically due to greatly enhanced continental weathering (Papineau *et al.*, 2007; Papineau *et al.*, 2009; Papineau, 2010), and perhaps matched the order of magnitude growth in the size of the continental land mass. Second, the stratification of the oceans would have been disrupted and increased mixing of waters with attendant upwelling along the coastal regions of the newly emergent continents would have also greatly enhanced nutrient availability throughout the oceans (Lowe, 1994; Cockell, 2000; Godderis and Veizer, 2000). Furthermore, to the extent the oceans may have cooled during the late Archean (Knauth, 2005; Lowe and Tice, 2007) this would have also created conditions more favorable to cyanobacteria while less so for thermophilic methanogens. Cooling would have also contributed to the reduction in ocean stratification and a deepening of the mixed layer (Cockell, 2000). Finally, another huge benefit to cyanobacteria and other ocean life of greatly increased continental input, reduced stratification, and enhanced upwelling would have been an increased presence in the surface ocean of various organic and inorganic UV-absorbing compounds and ions (Sagan, 1973; Cleaves and Miller, 1998; Garcia-Pichel, 1998; Cockell, 2000). These compounds and ions would have greatly attenuated the penetration through surface waters of any lethal UV not screened-out by the atmosphere and thus allowed cyanobacteria to inhabit more of the photic zone and increase in number.

The net effect of greatly improved oceanic nutrient availability, reduced lethal UV in surface waters, and possibly cooler temperatures would have been to trigger a large increase in the cyanobacterial population along with a concomitant increase in the $O_2$ source from oxygenic photosynthesis ($F_O$). On the other hand, the increasingly oxygenated state of the surface ocean between ~ 2.7 and 2.45 Gyr ago (Eigenbrode and Freeman, 2006; Kaufman *et al.*, 2007; Eigenbrode *et al.*, 2008; Godfrey and Falkowski, 2009; Kendall *et al.*, 2010), would have ultimately caused a complete collapse in the methanogen population inhabiting these waters. According to Kharecha *et al.* (2005), this segment of the methanogen population was especially productive; thus, its collapse would have led to a substantial decline in the $CH_4$ flux to the atmosphere. In addition, there would have been an increase in the surface ocean of $CH_4$ consumption by an expanding population of aerobic methanotrophs (Holland, 2009). This consumption would have included much of the $CH_4$ diffusing upwards from the deeper anoxic layer of the water column as well as any $CH_4$ still being produced in the upper layer. The relevant geochemical and microbiological conditions on Earth's surface during this period are modeled in Fig. 1.

As a result of these events the flux of $CH_4$ to the atmosphere significantly declined while the $O_2$ flux was increasing, and this would have helped trigger the rapid switch in the relative abundances of atmospheric $CH_4$ and $O_2$ at the time of the GOE. The net effect on atmospheric $O_2$ and $CH_4$ is shown in Fig. 2, and the sequence of events in the scenario presented here is consistent with one proposed by Zahnle *et al.* (2006): During ~ 2.6 Gyr to ~ 2.35 Gyr ago atmospheric $CH_4$ declined precipitously, the mass independent fraction (MIF) of sulfur disappeared, greenhouse warming collapsed, and atmospheric $O_2$ subsequently rose (Zahnle *et al.*, 2006; Domagal-Goldman *et al.*, 2008). The main difference between the model of Zahnle *et al.* (2006) and the one offered here is that instead of an increase in oceanic sulfate as the cause of the decline in the $CH_4$ (according to Habicht *et al.*, [2002] and Konhauser *et al.*, [2009] oceanic sulfate concentrations at this point were still very low), I propose that atmospheric $CH_4$ precipitously declined primarily because of a methanogen population crash in the surface ocean at the same time there was an expansion of the aerobic methanotroph population due to the increasingly oxygenated state of these waters during the period in question.

Aerobic methanotrophs probably started to become active at a level of dissolved $O_2$ as low as 50 nM (Ren *et al.*, 1997; Goldblatt *et al.*, 2006), which is well below the threshold of 2.5 µM (1% PAL) at which aerobic respiration is thought to usually begin (Knoll & Holland, 1995; Goldblatt *et al.*, 2006). By the time dissolved $O_2$ reached 5.7 µM, methanotrophic consumption of $CH_4$ would have achieved its maximum rate (Ren *et al.*, 1997), so further increases in dissolved $O_2$ would have not necessarily resulted in an increased amount of $O_2$ consumption, especially if the supply of $CH_4$ was dwindling due to the demise of methanogens in the surface ocean. Thus, $O_2$ could have continued to increase as $CH_4$ declined in these waters, even as aerobic methanotrophy could continue at its maximum rate.

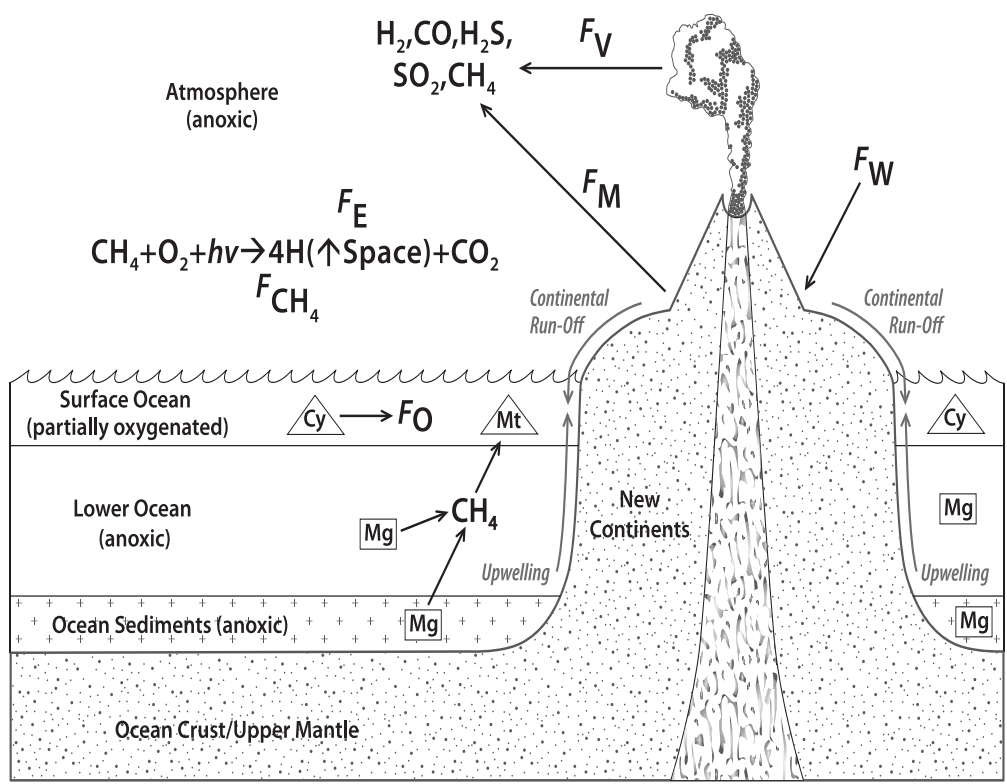

FIG. 1. Pertinent geochemical and microbiological conditions on Earth's surface just before and during the Paleoproterozoic Great Oxidation Event (GOE; ~ 2.6 – 2.4 Gyr ago). Various sinks for O2 are indicated by $F_V$ (reduced volcanic gases), $F_M$ (reduced metamorphic gases), $F_W$ (crustal weathering), and $F_{CH4}$ (methane oxidation). While a subaerial volcano is depicted, the reduced gases indicated would have also emanated from submarine volcanoes as well as serpentinization and hydrothermal vents in midocean ridge systems (Kasting, pers. communication). Methane oxidation and hydrogen escape to space are represented here as a summarized reaction according to Goldblatt et al. (2006). The source of $O_2$ is indicated by $F_O$ (oxygenic photosynthesis by cyanobacteria). Hydrogen escape to space from the photolysis of biogenic $CH_4$ is indicated by $F_E$. Symbols for microorganisms in the ocean important for the GOE are Cy = cyanobacteria, Mt = aerobic methanotrophs, and Mg = methanogens. Triangles enclosing Cy and Mt indicate growing populations of cyanobacteria and aerobic methanotrophs, while squares enclosing Mg indicate static populations of methanogens. Production of $CH_4$ by methanogens in the anoxic lower depths and sediments of the ocean and $CH_4$ consumption by aerobic methanotrophs in the partially oxygenated surface ocean are indicated.

An additional factor that probably contributed to the precipitous decline in atmospheric $CH_4$ before the GOE was a reduction from $\sim 400$ nM to $\sim 200$ nM of dissolved oceanic nickel, a key metal cofactor needed by several methanogen enzymes (Konhauser *et al.*, 2009). However, it seems that this reduction by half of oceanic nickel would have hardly, by itself, brought on a methanogen famine as argued by Konhauser *et al.* (2009). Indeed, 200 nM is still quite high compared to the modern level of dissolved oceanic nickel ($\sim 9$ nM; Drever, 1988), and according to Kida *et al.* (2001) methanogenic activity at 200 nM nickel is half that at 400 nM. Rather than a reduction by half in dissolved nickel, it would seem instead that the methanogen population in the surface ocean was decimated far more by the increasingly oxygenated state of these waters during the interval between $\sim 2.7 - 2.4$ Gyr ago, an increase which is supported by the results of Eigenbrode and Freeman (2006), Kaufman *et al.* (2007), and others.

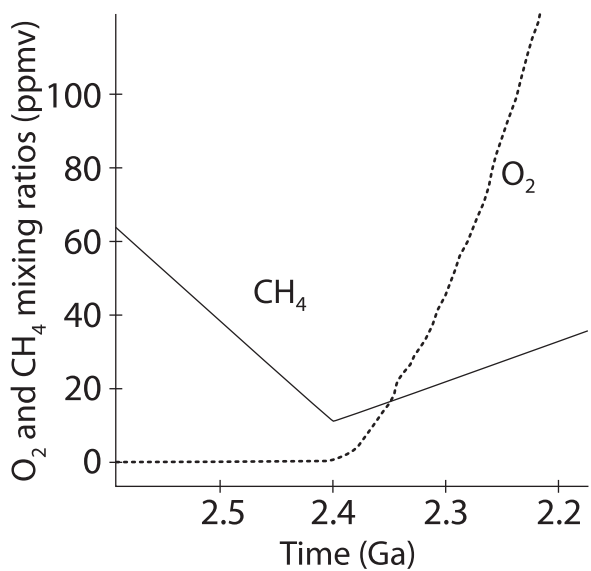

FIG. 2. Changes in the atmospheric mixing ratios (in parts per million by volume) of $O_2$ and $CH_4$ during the late Archaean and earliest Proterozoic as estimated by the biogeochemical modeling of Zahnle *et al.* (2006), and adapted here from Fig. 12 of that paper. According to Zahnle *et al.* (2006), as atmospheric $CH_4$ collapsed the mass-independent fractionation of stable sulfur isotopes disappeared When $CH_4$ reached a minimum at $\sim 2.4$ Ga (Gyr), greenhouse warming had declined sharply and a major ice age ensued. Shortly thereafter $O_2$ started rising to mildly oxic levels. Finally, after $\sim 2.4$ Ga, $CH_4$ started to increase due to the formation of a stratospheric ozone shield, and subsequently greenhouse warming was re-established.

### 3.3. *Further analysis of $O_2$ sources and sinks*

It should not go unnoticed that biogenic $CH_4$ apparently played a dual role in the rise of $O_2$ over geologic time, functioning as both an indirect source as well as a sink. Its function as an indirect $O_2$ source is represented by $F_E$ in equation (1), the escape of hydrogen into space as a result of the UV photolysis of biogenic $CH_4$ in the upper atmosphere (Catling *et al.*, 2001; Claire *et al.*, 2006). Conversely, it is parameterized as a sink by $F_{CH4}$ in equation (2) above, the destruction of $O_2$ via (biogenic) methane oxidation. According to Claire *et al.* (2006), "planetary oxidation via hydrogen escape is indirect" and "subtle." They argue that hydrogen escape became an $O_2$ source ($F_E$) by increasing the oxidation state of the crust and the metamorphic gases derived from the crust (Catling *et al.*, 2001; Claire *et al.*, 2006), thus decreasing the metamorphic gas sink represented by $F_M$ in equation (1). As discussed earlier, $F_M$ renders $F_E$ redundant and so $F_E$ does not appear in equation (2).

Because methanogens have probably existed on Earth since at least 3.46 Gyr ago (Ueno *et al.*, 2006) and atmospheric $CH_4$ concentrations were likely 1000 ppmv or higher during much of the Archaean (Kasting, 2005; Kasting and Ono, 2006), there was ample time (~ 1 Gyr) for hydrogen escape to oxidize the crust. This is because an atmospheric $CH_4$ concentration of 1000 ppmv would produce an effective flux of $O_2$ into the crust due to $CH_4$-induced hydrogen escape of about 7 X $10^{21}$ moles of $O_2$ over the course of 1 Gyr, an amount which is at least twice the modern continental crustal reservoir of excess $O_2$ (Catling *et al.*, 2001). Therefore, when atmospheric $CH_4$ started collapsing ~ 2.7 Gyr ago (Zahnle *et al.*, 2006), oxidation of the crust was likely to have been well advanced and the metamorphic gas sink ($F_M$) had probably declined substantially. To the extent volcanic gases by this time had also become less reducing (Kump and Barley, 2007; Gaillard *et al.*, 2011), $F_V$ was also smaller and thus the redox state of Earth's surface had become much more favorable for a rise in $O_2$. At this point I propose the most important factors that then triggered the GOE were the global expansion of the cyanobacterial population (Guo *et al.*, 2009) and $F_O$ during the burst of continent formation and rifting 2.7-2.4 Gyr ago, accompanied by a great decline in $F_{CH4}$. The combination of a greater source ($F_O$) and lower sinks ($F_M$, $F_V$, and $F_{CH4}$) caused the atmosphere to "flip" and become permanently oxygenated. The formation of stratospheric ozone after a low threshold of $O_2$ was reached in the lower atmosphere would have also contributed to the rapidity of the GOE by decreasing the rate of atmospheric methane oxidation (Claire *et al.*, 2006; Goldblatt *et al.*, 2006).

From this point forward, $O_2$ was probably always more abundant in the atmosphere than $CH_4$. Estimates of atmospheric $O_2$ concentration during and right after the GOE vary enormously, ranging from 0.1% PAL (200 ppmv) to ~15% PAL (30,000 ppmv [Holland, 2006; Holland, 2009]). However, even 0.1% PAL would have been sufficient to establish a stratospheric ozone layer which could block any lethal UV radiation (Kasting, 1987) that was not screened-out in other ways, and many life forms would have benefited from this. $CH_4$ would have also rebounded somewhat after the GOE because of the stratospheric ozone shield (Zahnle *et al.*, 2006), and it probably remained near 100-300 ppmv throughout most of the Proterozoic (Pavlov *et al.*, 2003; Kasting, 2005). Even if atmospheric $O_2$ stayed near the high end of the 0.1% - 15% PAL range until it rose even further in the Neoproterozoic, as long as $CH_4$ production was 10-20 times its modern value (as seems possible due to the remaining population of methanogens in the anoxic lower depths and sediments of the ocean), then the atmospheric $CH_4$ concentration could have persisted around 100-300 ppmv throughout the same period (Pavlov *et al.*, 2003; Kasting, 2005). The apparent stability of these conditions is reflected by the long period of stasis during the middle of the Proterozoic that has been called the "boring billion" (Holland, 2006). Until something caused the $CH_4$ flux to again decline and the $O_2$ flux to once more rise, apparently a state of equilibrium was able to prevail for a long time ($\sim$ 2.2 to $\sim$ 0.8 Gyr ago).

## 4. CONCLUSIONS AND FUTURE DIRECTIONS

It is concluded that the Paleoproterozoic GOE was in large part the result of biological processes – specifically, changes in the abundance and activity of cyanobacteria, methanogens, and methanotrophs – which occurred in conjunction with geophysical and geochemical processes such as hydrogen escape to space and changes in the redox state of volcanic gases. The great importance of biological factors in Earth's oxygenation has recently been highlighted by other investigators (e.g., see Hayes and Waldbauer, 2006; Konhauser *et al.*, 2009; Papineau, 2010). Of course, biological processes are often linked to and strongly influenced by geological events. For example, here it is argued that the population of cyanobacteria substantially increased between 2.7 and 2.4 Gyr ago mainly because a large burst of continent formation and rifting at that time led to a greatly enhanced level of nutrients and a reduced level of lethal UV in the surface ocean. Likewise, Konhauser *et al.* (2009) have attributed the decline in oceanic nickel during this period, which in turn contributed to the decline in $CH_4$ production by methanogens, to a cooling of the upper mantle.

The validity of the "biocentric" hypothesis presented here for Earth's oxygenation will be tested as additional geological and biological evidence is obtained. Accordingly, the following lines of investigation are proposed as ways of verifying or falsifying this hypothesis:

(1) The question of precisely when cyanobacteria and oxygenic photosynthesis arose continues to be a vexing one. As pointed out earlier, the range in estimates varies enormously from more than 3.7 Gyr ago (Rosing and Frei, 2004) to $\sim$ 2.5 Gyr ago (Kopp *et al.*, 2005). Hopefully more definitive evidence in the form of microbial biofilm and stromatolite fossils, microfossils, and molecular fossils (hydrocarbon biomarkers) will be forthcoming. Additional molecular evolutionary studies such as those of Xiong *et al.* (2000), Raymond *et al.* (2002), and Raymond *et al.* (2003) should also be helpful in more precisely timing the evolution of cyanobacteria. Finally, direct detection of $O_2$ in the geologic record such as the recent detection of $CH_4$ within 3.5 Gyr-old quartz mineral fluid inclusions (Ueno *et al.*, 2006) could also help to further constrain the temporal advent of cyanobacteria and oxygenic photosynthesis.

(2) Further studies of the distribution in the geologic record of various fossil types diagnostic for different microbes will help determine the expansion or contraction of cyanobacterial, methanogen, methanotroph, and other microbial populations through time, and thereby provide some idea of the pace at which they proliferated and enlarged their ranges or shrank to eventually inhabit relatively small niches. In addition to fossil data, further information on isotopic fractionation ratios in the geologic record for critical elements such as carbon, oxygen, nitrogen, and sulfur can also provide indications of the presence and size of various microbial populations, metabolic processes, and biogenic gas production over time (Catling and Kasting, 2007). Isotope fractionation data can also help elucidate the magnitude of the nitrogen cycle (as well as other elemental cycles), and thus provide more information about oceanic inventories of nitrogen and other nutrients over time. In the case of phosphorous, studies of phosphate-rich sedimentary rocks in the geologic record (e.g., Papineau, 2010) have already shed light on the oceanic phosphorous inventory over time. Taken together, further studies such as these will help to further elucidate to what extent cyanobacteria and other microbes may have experienced nutritional constraints on their proliferation.

(3) Trace metals are important for many metabolic processes and their abundance, isotopic composition, and distribution in the geologic record can be good proxies for the evolutionary and physiological status of various microbes (Anbar, 2008). This has been elegantly demonstrated in the case of oceanic nickel and the status of methanogens through geologic time by Konhauser *et al.*, (2009). Additional studies of this type on nickel and other bioessential metals such as iron, molybdenum, and vanadium will shed light, for example, on the abundance of diazotrophic cyanobacteria and other nitrogen-fixing microorganisms over time because of the requirement for iron and molybdenum or vanadium by the nitrogenase enzyme system (Anbar and Knoll, 2002; Raymond *et al.*, 2004).

(4) To more fully understand the rate at which aerobic microorganisms proliferated over geologic time and thus how quickly aerobic respiration became a significant $O_2$ sink, laboratory experiments similar to those of Zerkle *et al.* (2006) could be conducted which replicate Archean ocean waters with different temperatures, various levels of dissolved $O_2$ and NaCl, and nutrients including nitrogen, phosphorous, carbon and iron. Measuring the growth rates of a variety facultative aerobes in such simulated Archean ocean waters should reveal the levels of dissolved $O_2$ at which aerobic respiration would have began to confer a significant energetic advantage over fermentation (and other anaerobic metabolisms) and the exact magnitude of that advantage. Experiments such as these could also provide information about the rate of $O_2$ uptake under different ocean conditions with varying aerobic microbial population levels and thus make it possible to quantify the $O_2$ sink that would have existed under a given set of conditions.

(5) How quickly cyanobacteria may have caused the demise of methanogens in the Archean surface ocean could be tested by inoculating a laboratory simulation of the anoxic Archean ocean/atmosphere system with both types of microbes, and then monitoring the status of each population while measuring the amount of dissolved $O_2$ in the water.

In closing, determining the exact causes of Earth's oxygenation has other astrobiological ramifications. For example, detection of gaseous oxygen in the atmosphere of an Earth-like exoplanet would under most circumstances be considered diagnostic for the existence of life on that planet (Sagan *et al.*, 1993; Kasting, 2010; Léger *et al.*, 2011). To the extent that life on such an exoplanet is based on water, carbon, and biochemical and physiological processes similar to those on Earth, it might be expected that the oxygenation of this exoplanet (and others like it) resulted from the operation of oxygenic photosynthesis along with other biological and geological processes resembling those which have occurred on our planet.

## ACKNOWLEDGMENTS

The author is grateful to Jim Kasting, Lee Kump, Dominic Papineau, and Jason Raymond for reading and commenting on the manuscript as well as providing other very helpful suggestions. The author also thanks Tom Willard for assistance with Figures 1 and 2.

.